\documentclass[a4paper,aps,pre,twocolumn,floats,floatats,floatfix]{revtex4-1}
\usepackage[latin1]{inputenc}
\usepackage[brazil]{babel}
\usepackage{graphicx}  % Include figure files
\usepackage{dcolumn}   % Align table columns on decimal point
\usepackage{bm}        % bold math
\usepackage{natbib}
\usepackage{latexsym}
\usepackage{mathrsfs}
\usepackage{amssymb}
\usepackage{amsmath}
\usepackage{amscd}
\usepackage{color}
\usepackage{verbatim}
\usepackage{float}
\usepackage{lineno} %numeração de linhas
\begin{document}
\title{Aprendizagem significativa atrav\'es da modelagem computacional 
de sistemas f\'isicos \\
(Meaningful learning through computational modeling of physics systems)}
\author{H\'ercules A. Oliveira$^{1}$}
\email[E-mail address:~]{hercules@utfpr.edu.br}
\author{Caio V. Vieira$^{1}$}
\author{Tiago Kroetz$^{2}$}
%\thanks{}
\affiliation{$^{1}$Departamento Acadêmico de Matem\'atica, 
Universidade Tecnol\'ogica Federal do Paran\'a,
Ponta Grossa, PR, Brasil}
\affiliation{$^{2}$Departamento Acadêmico de F\'isica, 
Universidade Tecnol\'ogica Federal do Paran\'a,
Pato Branco, PR, Brasil}
\date{\today}

%
%%%%\linenumbers
%
\begin{abstract}
O avanço da ciência nos dias atuais depende fortemente do cálculo numérico, 
devido à possibilidade de se obter soluções através de simulações que seriam
inviáveis, ou mesmo impossíveis, de serem obtidas analiticamente. 
Nesse aspecto, torna-se de grande importância
dominar as ferramentas fundamentais de cálculo numérico aplicadas a alguma 
linguagem de programação, como por exemplo, da
programação em Fortran. 
Neste trabalho apresentamos alguns conceitos básicos de Fortran 77 e mostramos 
como estes podem ser aplicados à sistemas físicos, no intuito de oferecer um
primeiro contato, um guia, à esta linguagem de programação. 
Apresentamos a construção de rotinas que modelam o 
comportamento do sistema massa-mola, de funções trigonométricas e de um sistema 
mais complexo, o mapa padrão. 
Apresentamos uma discussão sobre a aprendizagem significativa 
envolvendo os conceitos físicos e a programação em Fortran.
\\
{\bf Palavras-chave: } Computação aplicada, Física computacional, Aprendizagem 
Significativa.
\\
\\
Nowadays the Science progress depends on the numerical calculus, due to the 
possibility of obtention of solutions using simulations which would be 
impracticable, or even impossible, to be analitically obtained.
In this aspect, it becomes important to dominate the fundamental tools of 
numerical calculus applied to some programming language, Fortran for example.
In this work we presented few basic concepts of the Fortran 77. 
We show how this concepts can be applied in the simulation and analysis of 
physical systems, trying to offer a first contact, a guide, with this 
programming language.
We presented the construction of routines that model the behavior of spring-mass
system, trigonometric functions and a more complex system, the standart map.
We presented a discussion about meaningful learning involving physical 
concepts and programming in Fortran.
\end{abstract}

%\keywords{Fortran, Cálculo numérico, Física computacional}

\maketitle

\section{Introdução}
\label{int1}
\par 
Diversas áreas da física e matemática utilizam o cálculo numérico para
visualizar a solução de equações simples ou até mesmo para verificar o 
comportamento de equações diferenciais lineares e não-lineares.
Para esses fins existem diversos programas que oferecem boa abordagem prática
como o Maple, Mathematica, Octave, Pyton, Fortran (costumeiramente escrito como 
FORTRAN), entre outros. 
Alguns dos programas citados, são exclusivos de programação numérica, como o 
FORTRAN.  
Outros dispõem de mais funções, como o cálculo analítico, caso do
Maple e do Mathematica. Muitos 
pesquisadores, em diversas áreas da física, utilizam atualmente a programação
em FORTRAN 77, FORTRAN 90 ou C$++$, principalmente por se tratarem de 
ferramentas gratuitas e amplamente difundidas nessas áreas. 

A linguagem de programação FORTRAN surgiu em 1953, com a proposta de J. Backus 
como a primeira linguagem de alto nível usada para programação de computadores
\cite{harry}.
Apenas em 1957 foi liberado o compilador para esta linguagem de programação.
O nome FORTRAN vem das palavras {\it FORmula TRANslation}, que indicam a 
transição de fórmulas matemáticas em liguagem de computadores. Diversos 
aprimoramentos surgiram nos anos seguintes e foram chamados de FORTRAN II 
em 1958, FORTRAN IV em 1961, FORTRAN 77 em 1977 e FORTRAN 90 em
1990. Outras versões intermediárias também foram lançadas, mas não citadas 
aqui, assim como a última versão de 2008. 
Apesar da evolução dos compiladores FORTRAN, e também do surgimento de outras 
linguagens de programação bastante funcionais, boa parte dos algoritmos 
utilizados em pesquisas de base continuam sendo implementados em FORTRAN 77.

O estudante de graduação em física, costumeiramente, 
é apresentado às necessidades 
de simulação computacional no decorrer de programas de iniciação científica. 
Com isto, geralmente nos primeiros passos deste aprendizado, o estudante
segue de forma autodidata e não muito eficiente. Muitas vezes, ele utiliza 
material didático inadequado ou ultrapassado.
O objetivo deste trabalho é propiciar uma aprendizagem mais significativa
oferecendo uma introdução \`a análise computacional 
de problemas físicos.
Busca-se despertar o interesse pelo estudo da física e da simulação
de sistemas físicos em linguagem FORTRAN. 
O presente estudo tem como norteador o caráter introdutório e aplicado
\`a física, sem a intenção de aprofundar em temas específicos de programação. 
Para maiores detalhes da linguagem de programação, 
como Constantes lógicas, Operadores lógicos, Extrutura condicional, 
o leitor deve procurar o 
livro da referência \cite{harry} ou manuais disponíveis na internet. 
O objetivo desse trabalho é mostrar que com conceitos básicos de programação
pode-se evoluir no entendimento de fenômenos físicos.
Nossos estudo é baseado na Linguagem de programação FORTRAN 77.

Este artigo está divido em sete seções. Na primeira seção (\ref{int1}) 
temos a introdução. Na seção (\ref{intro}) apresentaremos 
uma introdução aos conceitos da liguagem FORTRAN 77 e seus principais comandos.
Na seção (\ref{mecanica}) são apresentados alguns exemplos físicos de mecânica.
Na seção (\ref{nl}) apresentamos o rotor pulsado 
que dá origem ao Mapa Padrão ou Mapa de Chirikov-Taylor. Apresenta-se uma  
discussão sobre aprendizagem significativa na seção (\ref{as}). As conclusões
estão na seção (\ref{concl}) e por fim, os agradecimentos na seção (\ref{agra}).
\section{Introdução \`a Linguagem FORTRAN}
\label{intro}
O FORTRAN 77 (para simplificar, utilizaremos o termo ``Fortran'' para falar 
desta versão do FORTRAN)
apresenta algumas fases essenciais para a sua utilização em 
problemas físico-matemáticos de forma numérica, dadas pela sequência de 
etapas: 
1 - Modelagem do problema; 2 - Desenvolvimento de um algoritmo; 
3 - Implementação do algoritmo para a linguagem de programação; 
4 - Processamento do programa pelo computador e 5 - Análise dos resultados.

A etapa (1) trata da descrição detalhada e completa do problema a ser 
abordado pelo algoritmo, como a solução de uma integral ou de 
uma equação diferencial, explicitando quais os resultados esperados. 
A etapa (2) diz respeito \`a elaboração dos passos a serem adotados para 
chegar aos resultados. Na sequência, o algoritmo do problema é implementado
na linguagem de programação durante a etapa (3), ou seja, 
transcrever seu problema em comandos
ordenados para cálculo numérico. A quarta etapa se dá através da compilação e 
execução do programa (4), ou seja, a conversão do algoritmo em linguagem de 
máquina ou computador e do cálculo propriamente dito. 
Por fim, os resultados da simulação são analisados na etapa (5), 
extrai-se novas informações quantitativas e qualitativas do problema tratado 
e verifica-se a coerência destas informações com o conhecimento prévio do 
sistema.

Existem alguns comandos básicos da linguagem Fortran que devem 
ser apresentados inicialmente.
Vamos, de inicio, escrever os números inteiros de 1 até 10 em sequência. 
Primeiramente atribuímos um nome ao código, por exemplo ``sequencia'', 
sem acento.
O Fortran não faz distinção entre letras maiúsculas e minúsculas, por isso
é indiferente escrever ``sequencia'' ou ``SEQUENCIA'', no entanto os acentos 
não são caracteres aceitos pelo compilador
A linguagem exige que
as 6 primeiras casas, ou colunas, do editor de texto onde será escrito o código,
sejam destinadas a números com informações 
de repetição, por isso o programa deve começar na sétima casa. 

Na realidade o compilador recebe a informação das 80 primeira colunas do 
editor, sendo que destas, as colunas de 1 
\`a 5 são destinadas aos números que funcionam como rótulos dos comandos 
ou de declarações, quando usados. 
Caso o autor do código deseje inserir um comentário ao longo dos comandos, 
ele irá inserir o caractere ``c'' na primeira coluna do editor. Com isto, o 
compilador irá ignorar todas as linhas cuja primeira coluna contenha este 
caractere no momento em que for convertido em um programa executável.
A coluna 6 é utilizada para indicar a continuação da linha anterior e as 
colunas de 7 a 72 conterão as declarações do código. As colunas de 73 a 80 
são ignoradas pelo compilador.

Todos os comandos da linguagem são na língua Inglesa, por isso iniciamos com 
a palavra ``program'' seguido do nome do código a ser escrito.
Os espaços entre as palavras são ignorados quando o código for compilado.
Abaixo temos o começo do código com as colunas marcadas e o nome
do programa.
\begin{verbatim}
123456program sequencia
\end{verbatim} 
Completando o código temos
\begin{verbatim}
      program sequencia
      integer i
      do i=1,10
         write(*,*) i
      enddo
      end
\end{verbatim} 
Na primeira linha temos o início do código e o nome. Na segunda linha
a declaração da variável {\bf i} como sendo um número inteiro. Na terceira
o início da contagem com o comando {\bf do}, fazendo com que {\bf i} varie
de $1$ até $10$ em uma unidade inteira. Escreveremos os comandos em negrito 
para maior destaque e deixamos o espaço das seis colunas, sem que 
a contagem dos seis números apareçam, propositalmente para que o leitor sempre 
se atente a esse espaço utilizado pelo programa.
Na linha seguinte, pedimos para o programa escrever na tela os números 
através do 
comando {\bf write} e fechamos esta sequência com o comando {\bf enddo}. 
Por fim, na linha seguinte, fechamos o programa com um {\bf end}. 
Como vimos, utilizamos seis linhas para executar nosso cálculo. 
Cada linha deve conter um comando 
diferente conforme a necessidade. O resultado na tela será:
\begin{verbatim}
1
2
3
4
5
6
7
8
9
10
\end{verbatim}
O código deve ser salvo como um arquivo com extensão ``.f'', que indica ao
computador que se trata de um programa em Fortran. Exemplo: ``seq.f''.

Esse exemplo expressa a maneira de evoluir a variável em unidades. 
Caso se queira que a evolução de uma variável tenha um comportamento próximo 
do contínuo, basta somarmos números reais arbitrariamente pequenos antes de 
cada comando de impressão de seu valor na tela.
Isso pode ser expresso no código para encontrar o 
comportamento gráfico das funções Seno, Cosseno, Tangente, Secante, Cossecante 
e Cotangente, como mostra o código a seguir.
\begin{verbatim}
      program atrigo
      real*8 y1,y2,y3,y4,y5,y6,x,xf,passo
      open(1,file='sen.dat')
      open(2,file='cos.dat')
      open(3,file='tg.dat')
      open(4,file='cossec.dat')
      open(5,file='sec.dat')
      open(6,file='cotg.dat')
      pi=4.d0*datan(1.d0)
      x=0.01d0
      xf=2.d0*pi
      passo=0.05d0
 1    if(x.le.xf)then
         y1=dsin(x)
         y2=dcos(x)
         y3=dtan(x)
         y4=1.d0/y1
         y5=1.d0/y2
         y6=1.d0/y3
         write(1,*) x,y1
         write(2,*) x,y2
         write(3,*) x,y3
         write(4,*) x,y4
         write(5,*) x,y5
         write(6,*) x,y6
         x=x+passo
         goto 1
      endif
      end
\end{verbatim}
A primeira linha inicia ao código e lhe atribui um nome. 
Na segunda linha são definidas as variáveis reais, com precisão dupla de 15 
casas decimais (ex: 1,234567891234567), mas aparece na tela apenas 9 casas
(ex: 1,234567891). 
Da terceira até a oitava linha são abertos arquivos que 
irão armazenar os dados gerados pelo programa. Isto é feito com o comando 
{\bf open}, atribuindo-se um número para o arquivo, para controle do programa e 
um nome externo com a extensão {\bf .dat}.

Na linha nove está o comando {\bf datan(1.d0)}, que representa o arcotangente 
de $1$ ($arctg(1)$). Este comando multiplicado por $4$ fornece o valor 
numérico de $\pi$.
A letra "d" indica a utilização de dupla precisão.
 
Na sequência são atribuídos valores para o $x$ inicial $(x=0,01)$ e $x$ 
final $xf=2\pi$. A letra d seguida do número $0$ indicam que o número é real e
com dupla precisão, também indica que todas as casas decimais devem ser nulas 
até que se complete a precisão máxima da variável.
A separação decimal deve ser feita através do ponto 
na linguagem de computador (como em Inglês). A falta dessa descrição
({\bf d0}) traz imprecisões que podem ser acumulativas nas casas decimais 
subsequentes às seis primeiras.

Na linha 12 temos a definição do tamanho do passo considerado entre dois pontos
consecutivos. Na linha 13 iniciamos uma verificação, uma tomada de 
decisão. Aparecem os comandos {\bf if} (se) e {\bf then} (então) e o operador 
de relação {\bf le}.
O operador relacional {\bf le} determina se o valor da variável à esquerda do 
operador é menor ou igual ao valor da variável à direita e tem significado 
literal ``menor ou igual'' (less or equal). 
Existem ainda os operadores de relação {\bf lt} ``menor'', {\bf ge} 
``maior ou igual'' e {\bf gt} ``maior''. 
Nas linhas de 14 \`a 19 calculamos as funções 
trigonométricas utilizando os comandos já existentes no Fortran para o Seno, 
Cosseno e Tangente, que intrinsicamente são reais. 
Nas linha de 20 \`a 25 escrevemos o resultado do cálculo 
em seis arquivos. Na linha 26 fazemos a soma de números reais, acrescentando 
ao $x$ inicial $x=0,01$ mais $passo=0,05$. Na linha 27 utiliza-se o comando
{\bf goto 1} para que o programa retorne até o número $1$ na linha 13 para 
verificar a relação entre $x$ e $xf$, iniciando a pergunta se $x$ é menor ou 
igual que o $xf$. Caso a resposta seja positiva, o código
executa o cálculo novamente, caso contrário o programa avança para a linha
28 para o fechamento da pergunta com {\bf endif} e finalização do código
com {\bf end}, na linha 29. 

Na figura (1) e (2), temos o resultado do código atrigo, expresso através de 
gráficos. Estes foram gerados com os dados dos arquivos de saída definidos
no código.
Na figura (1) estão as
funções Cosseno $g(x)=cos(x)$ (na cor preta com pontos na forma de losângulos),
Seno $f(x)=sen(x)$ (na cor vermelha com pontos em forma de triângulos) e 
Tangente $h(x)=tg(x)$ (na cor azul com pontos na forma de círculos). 
No programa $h(x)=tg(x)$ é escrito em inglês $h(x)=tan(x)$. 
No gr\'afico expressamos $f(x)$, $g(x)$ e $h(x)$ 
simplesmente por $f(x)$. As funções estão 
delimitadas no eixo $x$ de $0$ até $2\pi$ e no eixo $y$ de $-1,5$ até $1,5$.
%
%fig 1
\begin{figure}[ht!]
\begin{center}
\includegraphics[height=6cm,clip]{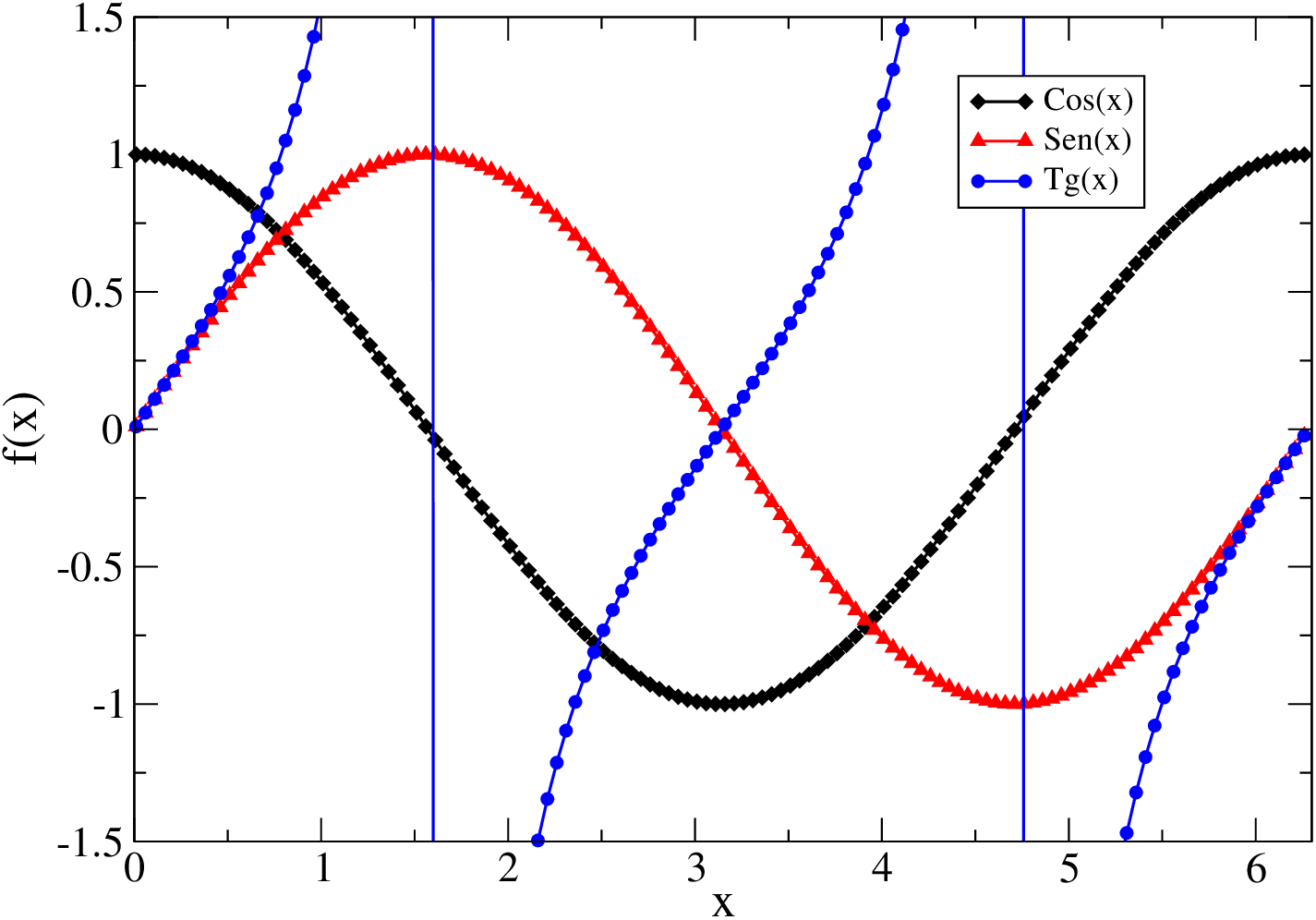}
\caption{\footnotesize{Gr\'afico das fun\c{c}\~oes 
Cosseno $g(x)=cos(x)$ (na cor preta com pontos na forma de losângulos),
Seno $f(x)=sen(x)$ (na cor vermelha com pontos em forma de triângulos) e 
Tangente $h(x)=tg(x)$ (na cor azul com pontos na forma de círculos).}}
\end{center}
\label{sen}
\end{figure} 
Na figura (2) temos o gráfico das funções 
Cossecante $g(x)=cossec(x)$ (na cor preta com pontos na forma de losângulos),
Cotangente $f(x)=cotg(x)$ (na cor vermelha com pontos em forma de triângulos) e 
Secante $h(x)=sec(x)$ (na cor azul com pontos na forma de círculos) 
pela variável $x$. 
%
%fig 2
\begin{figure}[ht!]
\begin{center}
\includegraphics[height=6cm,clip]{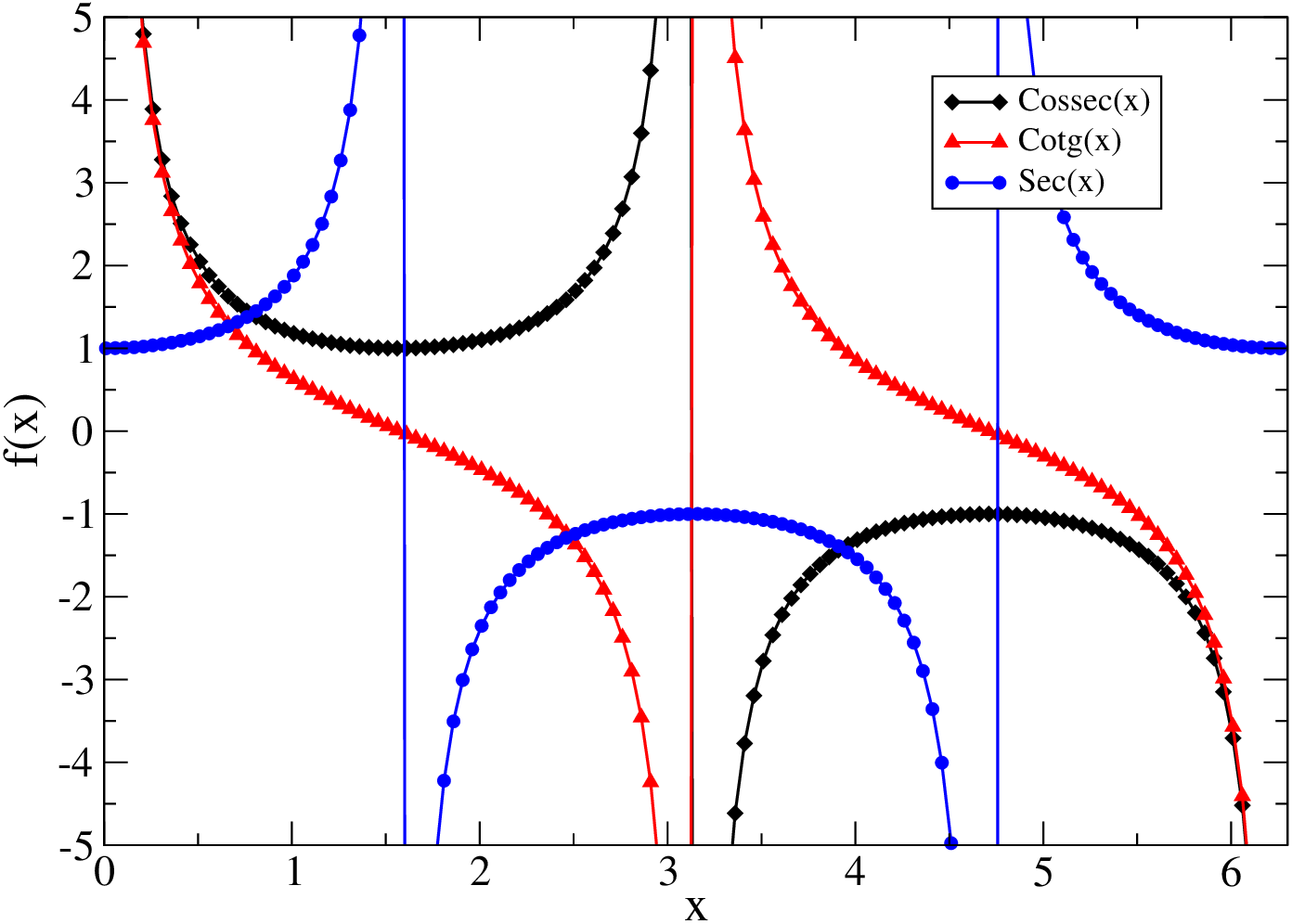}
\caption{\footnotesize{Gr\'afico das fun\c{c}\~oes 
Cossecante $g(x)=cossec(x)$ (na cor preta com pontos na forma de losângulos),
Cotangente $f(x)=cotg(x)$ (na cor vermelha com pontos em forma de triângulos) e 
Secante $h(x)=sec(x)$ (na cor azul com pontos na forma de círculos).}}
\end{center}
\label{cosec}
\end{figure}

\section{Aplicações em física: Mec\^anica}
\label{mecanica}
\subsection{Movimento de um corpo sob ação de força externa}
Come\c{c}aremos com a modelagem de uma part\'icula de massa $m$, 
movendo-se sob a a\c{c}\~ao
de uma for\c{c}a $F$ que gera uma acelera\c{c}\~ao $a$. Sabendo que a 
part\'icula possui posição inicial $x_{0}$ e velocidade inicial $v_{0}$, 
as equa\c{c}\~oes que regem o movimento desse corpo s\~ao \cite{hal}:
\begin{equation}
x=x_{0}+v_{0}t+\frac{at^{2}}{2}, \qquad v=v_{0}+at,
\label{mov1}
\end{equation}
para $(x)$ posição e $(v)$ velocidade em um instante de tempo $(t)$ e
\begin{equation}
F=ma,
\label{forca}
\end{equation}
para a rela\c{c}\~ao entre a força $F$, a massa e a aceleração. 
Esse tipo de movimento é chamado de 
movimento retilíneo uniformemente variado (MRUV) por manter uma aceleração 
constante fazendo com que a velocidade varie uniformemente \cite{hal}.

A interpreta\c{c}\~ao destas equa\c{c}\~oes podem ser mais facilmente obtidas
atrav\'es dos seus respectivos cálculos numéricos em Fortran,
como segue abaixo.
\begin{verbatim}
      program MRUV
      real*8 x0,v0,a,t,tfinal,passo,x,v
      real*8 m,F
      open(unit=1,file='posi.dat',
     $     status='unknown')
      open(unit=2,file='veloci.dat',
     $     status='unknown')
      x0=1.d0
      v0=1.d0
      m=1.d0
      F=10.d0
      a=F/m
      t=0.d0
      tfinal=10.d0
      passo=0.2d0
 1    if(t.lt.tfinal)then
         x=x0+v0*t+(a*(t**2))/2
         v=v0+a*t
         write(1,*) t,x
         write(2,*) t,v
         t=t+passo
         goto 1
      endif
      end
\end{verbatim} 

No código acima, temos o nome do programa na primeira linha. Na segunda e 
terceira linhas é feito a definição das variáveis e constantes quanto \`a sua 
natureza, sendo todas elas variáveis de números reais para este caso 
\cite{mdscha}. O comando {\bf real*8} denota essa informação. 

Na quarta linha temos o comando {\bf open}, que gera um arquivo de saída 
onde serão impressos os pontos com informação da posição da partícula a cada 
instante de tempo, seguido pela unidade a ser alocada dentro do programa
{\bf unit=1}. 
O nome do arquivo onde serão inseridos os dados é {\bf posi.dat}, 
onde ``dat'' é a extensão para arquivos com tabelas numéricas geralmente 
interpretados por programas de análise gráfica, como o Gnuplot ou XmGrace. 
Este arquivo será criado na pasta que
contém o código de execução. O {\bf status=unknown}, na linha abaixo, diz ao 
programa que ele precisa criar um arquivo ou escrever sobre esse arquivo já 
existente. Note que na linha do status existe o símbolo \$, que
representa a continuação da linha anterior.

Nas linhas 6 e 7 abrimos mais um arquivo para os dados referentes \`a velocidade
da partícula. Nas linhas de 8 a 11 inserimos os valores numéricos das 
constantes: posição inicial $(x_{0}=1)$, velocidade inicial $(v_{0}=1)$, massa da
partícula $(m=1)$ e força que atua sobre a mesma $(F=10)$. Numericamente não
existem unidades, mas sabemos que num sistema físico as grandezas acima são
expressas em Metro $(m)$, Metro por Segundo $(m/s)$, Quilograma $(kg)$ e 
Newton $(N)$, respectivamente. Na linha 12 encontramos escrito a equação 
(\ref{forca}) da força de maneira que se obtenha a aceleração da partícula.
Na linha 13 temos a definição do tempo inicial $({\bf t=0.d0})$ e do tempo 
final na linha 14 $({\bf tfinal=10.d0})$. Na linha 15 o tamanho do passo 
que determina o incremento de tempo entre dois pontos consecutivos do tempo 
$({\bf passo=0.2d0})$.

Na linha 16 excutamos uma pergunta utilizando os comandos {\bf if} 
(literalmente ``se'') e {\bf then} (``então''), na pergunta: ``Se t é menor
que o t final, então. 
A variável $t$ assume os valores correspondentes ao tempo transcorrido 
desde o início do movimento da partícula. 
Os comandos {\bf if} e {\bf then}
estão relacionados \`as tomadas de decisões em Lógica Matemática, 
onde simbolicamente tem-se 
$(a \to b)$, com significado ``se a, então b'' \cite{mdscha, rosen}.

Na linha 17 atribui-se à variável $x$ o valor da posição da partícula a cada 
instante de tempo $t$ utilizando-se da equação de movimento do M. R. U. V., das 
equações (\ref{mov1}), 
enquanto na linha seguinte atribui-se à variável $v$ o valor da velocidade a 
cada instante de tempo.
Nas linhas 19 e 20 escreve-se os valores das variáveis $t$ e $x$ para cada 
instante no arquivo {\bf 'posi.dat'}  e as variáveis $t$ e $v$ a cada instante 
no arquivo {\bf 'veloci.dat'}.

Acrescentamos \`a variável $t$ um incremento no valor de 
$0,2$ através da atribuição {\bf t=t+passo}. Em seguida, o código retorna 
a marcação $1$, com o comando
{\bf goto} ``vá para'', para repetir a pergunta na linha 16. 
Quando $t$ atingir o maior valor possível, mas menor que
$tfinal$, 
a sentença lógica entre parênteses, após o comando {\bf if}, deixará de ser 
verdadeira e a repetição dentro da estrutura deixará de ser executada.
Fecha-se essa pergunta com o comando
{\bf endif} (linha 23) e finalmente, fecha-se o código na linha 24 com o 
comando {\bf end} (``fim'').

Na figura (3) podemos ver o gráfico da posição da partícula em função dos 
instantes de tempo calculados pelo programa
para diferentes valores de força \`a qual ela está submetida. 
A curva em cor preta 
(linha preta e pontos em forma de losângulo) representa a posição em função 
do tempo para a força igual a $1$ 
($F=1$). A força $F=3$ está em vermelho com pontos triangulares. Para $F=5$ 
temos linhas na cor azul e pontos circulares. Para $F=7$ a linha é marrom e 
pontos em forma de mais ($+$) e para $F=10$, 
linha amarela com pontos em forma de asterísco.

As curvas da figura (3) mostram que a posição da partícula varia no 
tempo com um comportamento de uma função polinomial de grau 2, ou quadrático com
aceleração positiva (curva voltada para cima).
%fig. 3
\begin{figure}[ht!]
\begin{center}
\includegraphics[height=6cm,clip]{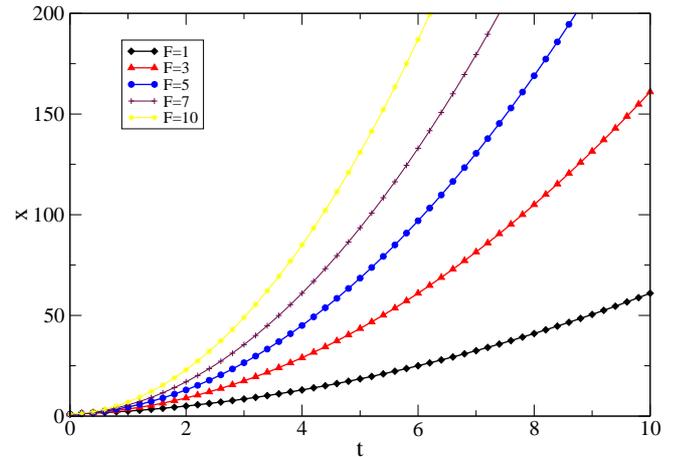}
\caption{\footnotesize{Gráfico da posição em função do  
tempo para diferentes valores de força. A curva em cor preta (linha preta e 
pontos em forma de losângulo) representa a posição 
em função do tempo para a força igual a $1$ ($F=1$). 
A força $F=3$ está em vermelho com pontos triangulares. Para $F=5$ 
temos linhas na cor azul e pontos circulares. Para $F=7$ a linha é marron e 
pontos em forma de mais ($+$) e para $F=10$, linha amarela com pontos em forma 
de asterísco.}}
\end{center}
\label{xport}
\end{figure} 

A figura (4) apresenta o gráfico da velocidade da partícula em função do 
tempo com diferentes valores de força. 
Esta figura apresenta o mesmo sistema de pontos e 
cores da figura (3) ($F=1$, Preta; $F=3$, Vermelha; 
$F=5$, Azul; $F=7$, Marron; $F=10$, Amarela). 

As retas da figura (4) mostram que a velocidade da partícula varia no 
tempo com um comportamento de uma função polinomial de grau 1, ou linear com
aceleração positiva (reta voltada para cima).
%
%fig. 4
\begin{figure}[ht!]
\begin{center}
\includegraphics[height=6cm,clip]{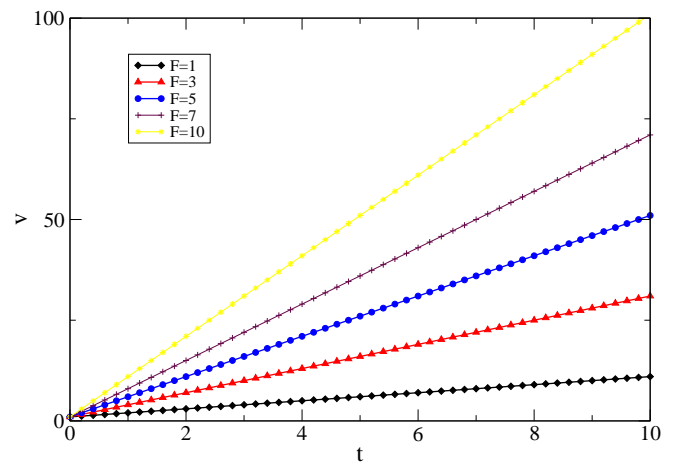}
\caption{\footnotesize{Gráfico da velocidade em função do 
tempo para diferentes valores de força. A curva em cor preta (linha preta e 
pontos em forma de losângulo) representa a posição 
em função do tempo para a força igual a $1$ ($F=1$). 
A força $F=3$ está em vermelho com pontos triangulares. Para $F=5$ 
temos linhas na cor azul e pontos circulares. Para $F=7$ a linha é marron e 
pontos em forma de mais ($+$) e para $F=10$, linha amarela com pontos em 
forma de asterísco.}}
\end{center}
\label{vport}
\end{figure} 
\subsection{Energia Mec\^anica}
\label{em}
Um dos conceitos mais fundamentais, quando inicia-se o estudo de física, é a 
Energia Mecânica. 
A conservação desta energia serve de base para a análise de inúmeros fenômenos 
e para a verificação da validade de muitas teo\-rias. 

O aluno que está iniciando seus estudos em física precisa de uma introdução 
detalhada sobre esse conceito, por isso vamos começar com o sistema mais 
simples apresentado nos livros textos básicos: o sistema massa-mola \cite{hal}. 
Este sistema é composto por um bloco, de massa $(m)$, 
preso a uma mola 
de constante el\'astica $(k)$ em uma superf\'icie sem atrito, como mostra a 
figura (5). Esse sistema tem energia cin\'etica $(K)$ (K maiúsculo, 
representando a palavra ``Kinetics'', que significa ``Cinética'' em inglês), 
atribuída ao movimento restrito do bloco, proporcional \`a sua 
massa e a sua velocidade ao quadrado. 
A energia potencial $(U)$ refere-se à compactação ou esticamento da mola.
Esta energia pode ser transferida para o bloco convertendo-se em movimento. 
A energia 
potencial do bloco, depende da constante da mola e 
da distância ao quadrado em relação \`a posição de equilíbrio 
estático ou de repouso. 
A constante $k$ (k minúsculo) é um valor específico que depende do tipo de 
material do qual a mola é feita e lhe dá maior ou menor resistência \`a 
compressão (esticamento).
Por isso, essa informação é essencial para saber quanta energia potencial 
o corpo irá adquirir e em quanto movimento irá se transformar.

%
%fig. 5
\begin{figure}[ht!]
\begin{center}
\includegraphics[height=4cm,clip]{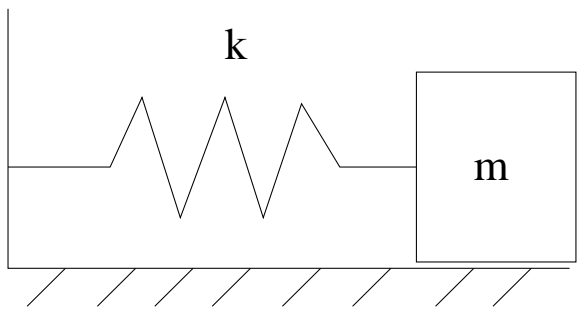}
\caption{\footnotesize{Sistema massa-mola. Bloco de massa $m$ e constante 
elástica $k$. Os traços em diagonal no solo representam o chão sem atrito.}}
\end{center}
\label{xport}
\end{figure} 

A energia cinética e a potencial podem ser escritas como
\begin{equation}
K=\frac{1}{2}mv^{2}, \qquad U=\frac{kx^{2}}{2}.
\label{eku}
\end{equation}

De forma simples, a soma das duas energias, $K$ e $U$, dão origem a chamada 
Energia Mecânica $(E)$. Aqui adotamos as nomenclaturas mais usuais para $K$, 
$U$ e $E$, outros símbolos podem ser encontrados dependendo da
bibliografia consultada, como por exemplo: 
$E_{c}$, para energia cinética ou $E_{p}$ para energia potencial e 
$E_{m}$, para energia mecânica.

Dessa forma, a energia mecânica é escrita matemáticamente como
\begin{equation}
E=\frac{1}{2}mv^{2} + \frac{kx^{2}}{2},
\label{em1}
\end{equation}
\begin{equation}
E=K+U.
\label{em2}
\end{equation}

Utilizando o gráfico destas três energias, percebe-se 
que as energias cinética, potencial elástica e mecânica possuem uma 
relação entre si.
A visualização desse gráfico auxilia muito no 
entendimento do conceito das três energias e faz com que o aluno crie conexões 
entre o gráfico, o conceito e as equações. Essa abordagem é parte integrante da 
aprendizagem no que diz respeito ao desenvolvimento cognitivo em física.

Outro mecanismo de fixação do conceito e das equações é a construção do 
código que simula o sistema. 
Assim o aluno entende a lógica que envolve o conceito 
físico. Tal código em Fortran é apresentado a seguir.
\begin{verbatim}
      program mola
      real*8 x,a,U,K,E,xfinal,passo
      open(unit=1,file='u1.dat',
     $     status='unknown')
      open(unit=2,file='k1.dat',
     $     status='unknown')
      open(unit=3,file='e1.dat',
     $     status='unknown')
      x=-3.d0
      xfinal=3.2d0
      passo=0.2d0
      a=5.d0
      E=22.5d0
 1    if(x.lt.xfinal)then
         U=(a*(x**2))/2.d0
         K=E-U
         write(1,*) x,U
         write(2,*) x,K
         write(3,*) x,E
         x=x+passo
         goto 1
      endif
      end
\end{verbatim}
%
%fig. 6
\begin{figure}[ht!]
\begin{center}
\includegraphics[height=6cm,clip]{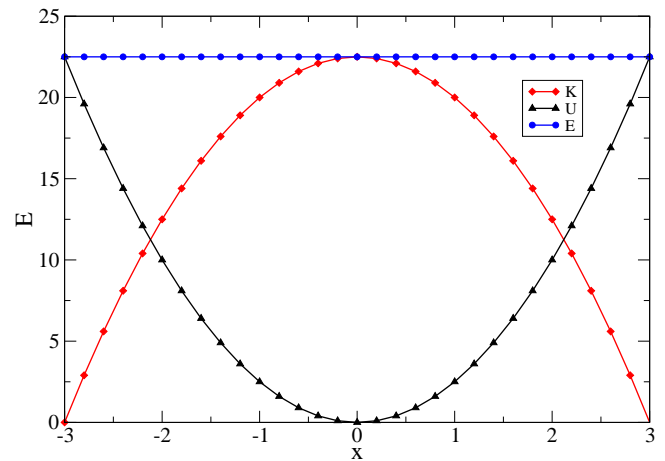}
\caption{\footnotesize{Gr\'afico das energias 
cinética (em vermelho com pontos em losângulo),
mecânica (em azul com com pontos circulares) e 
potencial (em preto com pontos triangulares), 
relacionadas ao movimento harmônico simples.}}
\end{center}
\label{energiacpm}
\end{figure} 

Apresentam-se os mesmos comandos básicos para início de código, como 
feito nos códigos anteriores. Atribui-se nome ao programa (linha 1), 
define-se as variáveis reais (linha 2), nas linhas de 3 à 8 abrimos 
arquivos externos que receberão os valores de energia potencial, 
energia cinética e mecânica respectivamente.
A variável principal nesse programa é a posição $(x)$. Estabelecemos uma posição
inicial $x=-3$ e final $xfinal=3,2$ com passo de variação de $0,2$. A Energia 
mecânica total foi fixada em $E=22,5$ e a constante da mola tendo valor $k=5$ 
(``k'' foi chamado de ``a'' no código, pois já utilizou-se ``K'' para denotar 
a energia cinética).
Nota-se que não é necessário saber a velocidade $(v)$ da massa $(m)$, 
pois a energia cinética é obtida através da diferença entre a energia mecânica 
e a potencial em cada posição.
A conservação de energia pode ser utilizada por quê o sistema é conservativo 
\cite{hal, goldstein}. 
Esse procedimento é feito através da equação (\ref{em2}), na 
linha 16 do código. Dessa forma, temos da 
energia mecânica (linha 13) constante, o cálculo da energia potencial na linha 
15 e a energia cinética na linha 16.

A variação dos valores assumidos pela variável $x$ se dá na linha 20, 
enquanto a 
verificação de que ela ainda possui valor inferior ao máximo estabelecido 
($xfinal$) está expresso na linha 14.
Para a primeira verificação, o programa compara os valores de $x=-3$
com $xfinal=3,2$. Então ({\bf then}) ele continua e realiza todos os cálculos
solicitados chegando \`a linha 20 que adiciona $0,2$ \`a posição $x$ e passa ao 
próximo comando {\bf goto 1} que determina a execução do programa a retornar 
ao número 1 anterior fixado na 
linha 14. Dessa forma, criamos uma conta cíclica (um ``loop'') até atingirmos o
valor suficientemente alto para a variável de posição $x$. 

A figura (6) mostra a representação gráfica resultante do código mola.
São apresentadas as três energias do sistema em função da posição da massa. 
A energia potencial (linha preta com pontos triangulares) começa com um valor 
total de $22,5$, na posição $x=-3$ e diminui até atingir o valor mínimo $U=0$
 em $x=0$. Ela retorna a aumentar até o valor inicial em $x=3$. O comportamento 
da energia cinética (linha vermelha com pontos em forma de losângulos) é o 
inverso. Esta energia inicia com valor $K=0$ e aumenta até o $K=22,5$ em $x=0$,
 exatamente na posição em que a energia potencial é zero. 
Com o aumento dos valores da variável $x$, $K$ 
dimunui até atingir zero novamente. Esse comportamento ocorre por causa da 
massa presa à mola. Quando comprimimos (esticamos) a mola (entende-se que a 
massa está presa à mola) o sistema tem energia potencial máxima. 
Quando liberamos a massa, essa energia é transformada 
em movimento e a energia cinética apresenta valores não nulos.
Como a energia mecânica (linha azul 
com pontos circulares) é a soma das duas outras energias ela se mantém constante
durante todo o tempo, como é de se esperar num sistema conservativo.

\section{Din\^amica N\~ao-Linear}
\label{nl}
\subsection{O Rotor Pulsado}
A Teoria do Caos apresenta diversos fenômenos em física e matemática, 
que já são amplamente estudados por diversos pesquisadores, 
mas que ainda deixam leigos muito intrigados. 
A dinâmica caótica está presente em problemas na área chamada de
Dinâmica Não-linear ou de Sistemas Dinâmicos Caóticos. 
Certos fenômenos são bem comuns, como os Fractais, Atratores caóticos 
(que possuem relação com o que se conhece como efeito borboleta), 
Órbitas periódicas e caóticas e Expoentes de Lyapunov 
\cite{her2, goldstein, monteiro, lichtenberg}. 
Para introduzir esse tema na linguagem Fortran apresentaremos o Mapa Padrão, 
ou Mapa de Chirikov-Taylor \cite{Chirikov}, que
é resultado da discretização das equações de movimento do Rotor pulsado. 

O rotor pulsado é um sistema dinâmico composto por uma haste que pode girar em torno de um eixo após 
sofrer ação de uma força.
Consideremos uma haste, de comprimento $L$, com uma extremidade fixa de modo 
que a mesma possa fazer o movimento de rotação sobre esse eixo, como 
apresentado na figura (7). 
%
%fig. 7
\begin{figure}[ht!]
\begin{center}
\includegraphics[height=6cm,clip]{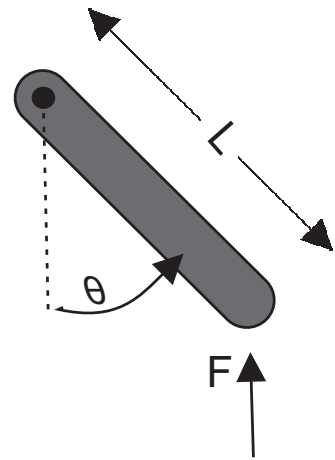}
\caption{\footnotesize{Haste com eixo de rotação (Rotor pulsado) com 
comprimento $L$ sob ação de um for\c{c}a $F$, que a faz rotacionar 
angularmente em $\theta$.}}
\end{center}
\label{KR}
\end{figure} 

Sobre esse sistema atua uma for\c{c}a $F$ que depende de um campo 
gravitacional $g$. A força oscila com o tempo $(t)$ e pode ser escrita como um 
somatório de pulsos da forma
\begin{equation} 
F(t)=f\sum_{n=-\infty}^{\infty}{\delta (t-nT)}.
\label{fi}
\end{equation}
Nesta equação, $T$ \'e o per\'iodo de oscila\c{c}\~ao, $\delta$ \'e a 
fun\c{c}\~ao delta de Dirac e $f$ é a intensidade da for\c{c}a.

O potencial da haste \'e dado por 
$V=-\int{F(t)dx}= Fx = FLcos(\theta)$. Fazendo algumas manipula\c{c}\~oes
alg\'ebricas encontramos
\begin{equation}
V=I\omega_{0} ^{2} cos(\theta)\sum_{n=-\infty}^{\infty}{\delta (t-nT)}.
\label{v}
\end{equation}
onde $\omega_{0} ^{2} =g/l$ \'e a frequência natural de oscila\c{c}\~ao da 
haste e $I=mL^{2}$ é o momento de inércia de uma haste.
A energia cin\'etica da haste \'e uma energia de rota\c{c}\~ao pura e podemos
expressar como
\begin{equation}
T=\frac{1}{2}I\omega ^{2}= \frac{p^{2}}{2I},
\label{t}
\end{equation}
com $\omega = \frac{d\theta}{dt}$ 
a velocidade angular ou em termos do momento linear $(p)$ \cite{hal, goldstein}.

A Hamiltoniana do sistema \'e escrita pela soma da energia cin\'etica com a 
energia potencial \cite{lemos, her}, dada por
\begin{equation}
H(p, \theta, t) = \frac{p^{2}}{2I} + 
I\omega_{0} ^{2}cos(\theta)\sum_{n=-\infty}^{\infty}{\delta (t-nT)}.
\label{hamil}
\end{equation}
As equa\c{c}\~oes de movimento s\~ao dadas pelas equações de Hamilton 
\begin{eqnarray}
\dot{\theta} &=& \frac{dH}{d p} = \frac{p}{I} 
\label{tetap}
\\
\dot{p} &=& - \frac{dH}{d \theta }= I\omega_{0} ^{2}sen(\theta) 
\sum_{n=-\infty}^{\infty}{\delta (t-nT)}
\label{pp}
\end{eqnarray}
O ponto sobre as vari\'aveis significa a derivada temporal 
$\dot{\theta} = \frac{d\theta}{d t}$.

Os pulsos de for\c{c}a ocorrem nos instantes $t=nT$, onde n é o enésimo 
impacto que ocorre a cada intervalo de tempo $T$.
Integrando a equação (\ref{pp}) durante o intervalo de tempo entre impactos 
consecutivos
\begin{eqnarray}
&&\int_{nT}^{(n+1)T}{\dot{p} dt} = \int_{nT}^{(n+1)T}{\frac{dp}{dt} dt}
\nonumber
\\
&=&
\int_{nT}^{(n+1)T}{I\omega_{0} ^{2}sen(\theta) 
\sum_{n=-\infty}^{\infty}{\delta (t-nT)}dt}
\label{intp1}
\end{eqnarray}
Considerando um tempo infinitesimal, pode-se fazer a aproximação 
$sen(\theta) \sum_{n=-\infty}^{\infty}{\delta (t-nT)} =sen(\theta _{n})$, 
onde $\theta _{n}$ é o valor da variável angular do rotor no instante do 
enésimo impacto. Com isso a variável $\theta _{n}$ não depende do tempo e a
integral pode ser escrita como:
\begin{eqnarray}
\int_{nT}^{(n+1)T}{dp}=p_{n+1}-p_{n} &=&
\int_{nT}^{(n+1)T}{I\omega_{0} ^{2}sen(\theta) dt}
\nonumber
\\
&=& I\omega_{0} ^{2}sen(\theta _{n}) T,
\label{intp2}
\end{eqnarray}
onde $T=t_{n+1}-t_{n}$ \'e o per\'iodo de oscila\c{c}\~ao da for\c{c}a
e $p_{n}$ é o valor do mometo angular do rotor no instante do enésimo impacto.
Com isso, temos
\begin{equation}
p_{n+1} = p_{n} + I\omega_{0} ^{2}Tsen(\theta _{n}).
\label{pn1}
\end{equation}

Para obtermos o valor da variável angular discretizada em cada impacto,
integramos a equação (\ref{tetap}) e obtemos
\begin{eqnarray}
\int_{nT}^{(n+1)T}{\dot{\theta} dt} &=& \int_{nT}^{(n+1)T}{\frac{d\theta}{dt} dt}
\\
&=&
\int_{(n)T}^{(n+1)T}{d\theta}=\int_{nT}^{(n+1)T}{\frac{p(t)}{I}dt}.
\label{intt1}
\end{eqnarray}
Resolvendo a integral para $\theta$, temos
\begin{equation}
\theta _{n+1}-\theta _{n} = \int_{nT}^{(n+1)T}{\frac{ p_{n+1} }{I}dt}
=\frac{p_{n+1} }{I}T.
\label{teta1}
\end{equation}
Resultando em
\begin{equation}
\theta _{n+1} = \theta _{n} +  \frac{T}{I} p_{n+1}.
\label{tetaf}
\end{equation}
Escolhendo exatamente o per\'iodo $T=I$ e fazendo uma mudança de variáveis 
$K=(I\omega)^{2}$, obtemos o mapa padr\~ao descrito através das 
equações discretizadas por:
\begin{eqnarray}
p_{n+1} &=& p_{n}+Ksen(\theta _{n}) 
\label{pmp}
\\
\theta _{n+1} &=& \theta _{n} +p_{n+1}
\label{tmp}
\end{eqnarray}
ou por
\begin{eqnarray}
p_{n+1} &=& p_{n}+Ksen(\theta _{n})  \hspace{0.5cm} mod \hspace{0.2cm} 2\pi
\label{pmp2p}
\\
\theta _{n+1} &=& \theta _{n} +p_{n+1} \hspace{1.3cm}  mod \hspace{0.2cm} 2\pi
\label{tmp2p}
\end{eqnarray}
Escolhemos novamente a letra K para mostrar exatamente como o mapa padrão é
encontrado em livros e artigos, mas não deve-se confundir com a constante da
mola ou energia cinética. A constante K aqui é adimensional sem ligação
com as grandezas físicas anteriores.

O $mod \hspace{0.2cm} 2\pi$ faz com que as solu\c{c}\~oes fiquem dentro de um 
espa\c{c}o de fases modulado em $2\pi$ e o retrato de fases aparece nos 
intervalos de $0$ à $1$. 
Em diversos estudos pode-se encontrar a posição angular $\theta$ 
expressa pela variável $x$. Neste caso temos as dimens\~oes 
como $(x, p)$ e as equações na forma
\begin{eqnarray}
p_{n+1} &=& p_{n}+Ksen(x _{n}) 
\label{px}
\\
x _{n+1} &=& x _{n} +p_{n+1}
\label{xx}
\end{eqnarray}
As variáveis são escritas com índices $x_{n}$, $p_{n}$ para indicarem que o tempo
utilizado na evolução do sistema é discreto, de forma que $n$ assume
valores inteiros $n= 1, 2, 3, \cdots , j$. 
Já as variáveis $x_{n}$ e $p_{n}$ assumem valores reais, 
como $x_{2}=0,34$ e $x_{5}=0,62$, por exemplo.
\subsection{Retrato de fase do Mapa Padr\~ao}
Modelando o Mapa Padrão em Fortran podemos escrever o seu retrato de fase, que
é o armazenamento das variáveis $x$ e $p$ com o passar 
do número de impactos (ou iterações das equações). 
Constroi-se um gráfico com $x_{n+1}$ e $p_{n+1}$ para verificar como a 
dinâmica do sistema se comporta para cada valor do parâmetro $K$. 

Inicia-se o código com um nome na primeira linha e na segunda linha 
utilizamos o comando que atribui valor real a todas às variáveis que começam 
com as letras
de a até h e de o até z, com exceção das letras i, j, k, l, m, n,
que eventualmente podem ser designadas como variáveis inteiras.
Essa definição é apresentada na linha 6.
A linha 3 define que os valores de $n$ e $n1$ podem chegar a valores iguais a
$20.000$ e $20.001$, por escolha arbitrária. Na linha 4 é definido que as 
variáveis $p$ e $q$, reais, serão representadas por vetores de dimensão iguais 
à $n1$. Pode-se determinar que os valores de $n$ e $n1$ sejam iguais ao 
tamanho do vetor a ser utilizado.
\begin{verbatim}
program mapapadrao
implicit real*8(a-h,o-z)
parameter(n=20000,n1=20001)
dimension p(n1),q(n1)
EXTERNAL RAN1
integer i
open(1,"pmapa.dat")
pi = 4.d0*datan(1.d0)
a=0.18d0
do j=1,300
   q(1)= (ran1(idum))
   p(1)= (ran1(idum))+0.1d0
     do i=1,15000
        p(i+1)=p(i)+a*sin(2.d0*pi*q(i))
        q(i+1)=q(i)+p(i+1)
        p(i)=p(i+1)
        q(i)=q(i+1)
        if(i.gt.9000)then
          if(p(i).gt.0.d0)then
            if(q(i).gt.0.d0)then
              write(1,*) mod(q(i),1),
             $           mod(p(i),1)
            endif
          endif
        endif
     enddo
enddo
end

FUNCTION ran1(idum)
INTEGER idum,IA,IM,IQ,IR,NTAB,NDIV
REAL*8 ran1,AM,EPS,RNMX
PARAMETER (IA=16807,IM=2147483647,AM=1.d0/IM,
$     IQ=127773,IR=2836,
      *NTAB=32,NDIV=1+(IM-1)/NTAB,EPS=1.2d-7,
$     RNMX=1.d0-EPS)
      INTEGER j,k,iv(NTAB),iy
      SAVE iv,iy
      DATA iv /NTAB*0/, iy /0/
      if (idum.le.0.or.iy.eq.0) then
        idum=max(-idum,1)
        do 11 j=NTAB+8,1,-1
          k=idum/IQ
          idum=IA*(idum-k*IQ)-IR*k
          if (idum.lt.0) idum=idum+IM
          if (j.le.NTAB) iv(j)=idum
11      continue
        iy=iv(1)
      endif
      k=idum/IQ
      idum=IA*(idum-k*IQ)-IR*k
      if (idum.lt.0) idum=idum+IM
      j=1+iy/NDIV
      iy=iv(j)
      iv(j)=idum
      ran1=min(AM*iy,RNMX)
      return
      END
\end{verbatim}
%%%%%%%%%%%%555
%
O comando {\bf EXTERNAL RAN1} traz para o código principar um código 
subjacente, que realiza o cálculo apenas em determinado local,
quando solicitado, sem ocupar tempo de programação desnecessária. 
Esse tipo de sub-código é chamado de ``sub-rotina'' (ou {\it subroutine} em 
inglês) e aparece na sequência do final do código principal. 
Neste caso, a sub-rotina é a {\bf RAN1} que é utilizada nas 
linhas 11 e 12 para gerar números aleatórios às variáveis $x_{1}$ e 
$p_{1}$, definidas como $q(1)$ e $p(1)$ no código principal. Essa sub-rotina 
pode ser encontrada com melhor descrição na referência \cite{nreci}.

Na linha 8 está o comando que fornece o valor numérico de $\pi$, 
como mencionado anteriormente.
O parâmetro $K$, presente na equação (\ref{px}), é definido na linha 9, 
com a letra $a$ no código por dois motivos, um por simplicidade outro para 
que ele seja usado como um número real, 
já que definimos inicialmente que a letra $k$ seria uma variável inteira. 

Depois de escolher aleatóriamente os valores de $x_{1}$ e $p_{1}$ iniciamos a 
iteração do mapa através dos comando {\bf do} com a contagem de $i=1$ até 
$i=15.000$. O programa insere os primeiros valores de $x$ e $p$ nas  
equações (\ref{px}) e (\ref{xx}) obtendo os valores das variáveis 
($x_{2}$ e $p_{2}$) num tempo seguinte. 
Neste caso, o tempo é discreto, ou seja, $i$
varia em apenas uma unidade inteira $(i = 1, 2, 3, \cdots, 15.000)$. Os dez
primeiros valores do tempo discreto $i$ e das variáveis reais $x$ e $p$ podem 
ser vistos na tabela (\ref{tab1}). Pode-se notar que a iteração do mapa nada 
mais é do que a utilização de duas funções compostas, feito nas linhas de 13 
\`a 17.

\begin{table}[h]
\centering
\caption{Tabela de valores para $x_{i}$ e $p_{i}$ em função de $i$ com nove 
casas decimais.}
\begin{tabular}{|c|c|c|c|c|}
\hline
i & $x_{i}$ & $x_{i+1}$ & $p_{i}$ & $p_{i+1}$ \\
\hline
1 & 0,415999357 & 0,698617052 & 0,191964891 & 0,282617695 \\
\hline
2 & 0,698617052 & 0,810534358 & 0,282617695 & 0,111917306 \\
\hline
3 & 0,810534358 & 0,755315313 & 0,111917306 & -0,055219045 \\
\hline
4 & 0.755315313 & 0.520196642 & -0.055219045 & -0.235118671 \\
 \hline 
5 &  0.520196642 &  0.262297363 & -0.235118671 & -0.257899279 \\
\hline 
6 &  0.262297363 &  0.18386104 & -0.257899279 & -0.0784363231 \\
\hline 
7 &  0.18386104  & 0.270104744 & -0.0784363231  & 0.0862437042 \\
\hline 
8 &  0.270104744 &  0.534914208 &  0.0862437042  & 0.264809464 \\
\hline 
9 &  0.534914208 &  0.760552586  & 0.264809464 &  0.225638378 \\
\hline 
10 &  0.760552586  & 0.806586478 &  0.225638378 &  0.0460338913 \\
\hline 
\end{tabular}
\label{tab1}
\end{table}
%fig. 8
\begin{figure}[ht!]
\begin{center}
\includegraphics[height=6cm,clip]{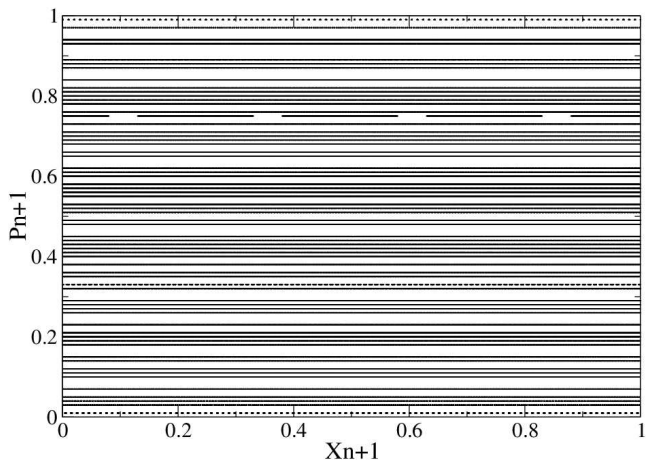}
\caption{\footnotesize{Retrato de fase do mapa padrão ($p_{n+1}$; $x_{n+1}$) 
para $K=0$.}}
\end{center}
\label{rf0}
\end{figure} 
O sistema fornece grande oscilação nos primeiros valores obtidos, estabilizando 
depois de certo tempo para valores de $K$ onde o mapa é periódico. 
Essa oscilação é chamada de Transiente e tenta-se 
eliminá-lo não considerando certos valores iniciais. Nesse código 
desconsideramos os valores de $x$ e $p$ para o tempo de $i=1$ até $i=9.000$. 
Registramos os valores das variáveis para tempos de $i=9.001$ até $i=15.000$. 
Isso pode ser visto na linha 18, com a relação {\bf if}, que se fecha na linha
25 com {\bf endif}. Nas linhas 19 e 20, também com {\bf if}, selecionamos
apenas os valores de $p$ e $q$ positivos. 
Esse procedimento é necessário para não aparecer o mesmo gráfico replicado da 
parte positiva na parte negativa. A reflexão, da parte positiva na parte 
negativa do retrato de fases é uma propriedade do mapa padrão.

Nas linhas 21 e 22 escrevemos as variáveis $q_{i}$ e $p_{i}$ aplicando o 
{\bf mod(q,1)} \cite{santos} para que seus valores mantenham-se entre $0$ e 
$1$. Após esse procedimento,
fecha-se o programa com os {\bf endif} e {\bf enddo} nas linhas seguintes. 

Deixando uma linha em branco para iniciar a sub-rotina de números aleatórios, 
que não entraremos em detalhes aqui. Sugerimos ao leitor procurar a 
refer\^encia \cite{nreci} para maiores detalhes desse código.
%%%%%%%%
%
\begin{figure}[ht!]
\begin{center}
\includegraphics[height=6cm,clip]{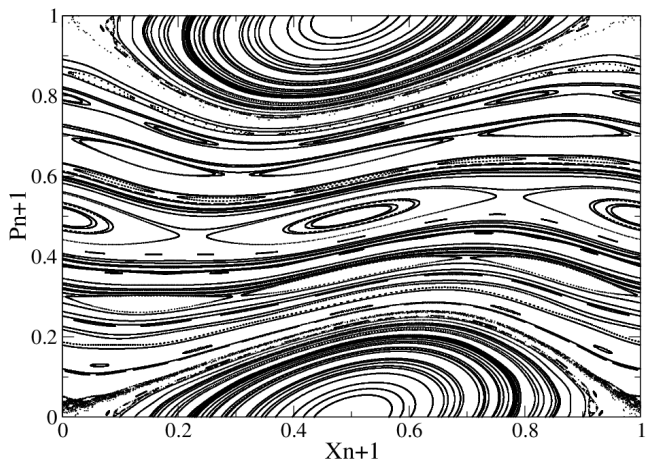}
\caption{\footnotesize{Retrato de fase do mapa padrão para $K=0,1$.}}
\end{center}
\label{rf01}
\end{figure} 

Nas figuras (8), (9), (10), (11) e 
(12) temos o resultados do programa do mapa padrão detalhado até agora.
A figura (8) apresenta o retrato de fase \cite{monteiro, her2}, $p_{n+1}$ por
$x_{n+1}$ (n ou i no código), para o Mapa Padrão com condições iniciais 
$(x_{1}, p_{1})$ aleatórias e parâmetro $K$ sendo igual a $0$.
Para esse valor de $K$ o sistema é integrável (completamente regular)
\cite{monteiro, her2} e apresenta linhas horizontais. 
O sistema não apresenta caos e só pode haver 
valores iguais de $p_{n+1}$ para diferentes $x_{n+1}$.
%%%%%%%%%
\begin{figure}[ht!]
\begin{center}
\includegraphics[height=6cm,clip]{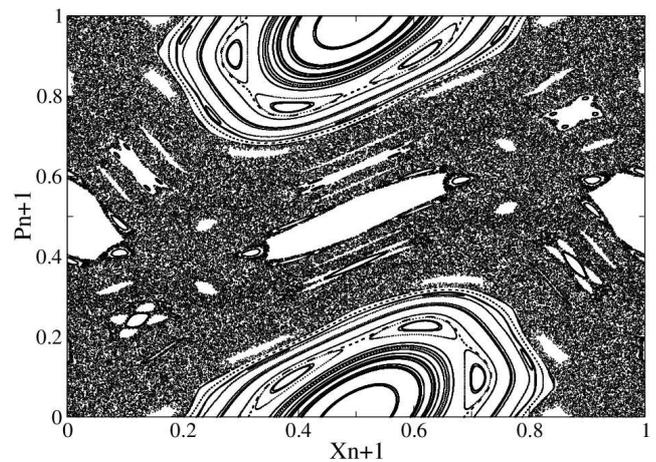}
\caption{\footnotesize{Retrato de fase do mapa padrão para $K=0,2$.}}
\end{center}
\label{rf02}
\end{figure} 
%
%%%%%%%%%5
\begin{figure}[ht!]
\begin{center}
\includegraphics[height=6cm,clip]{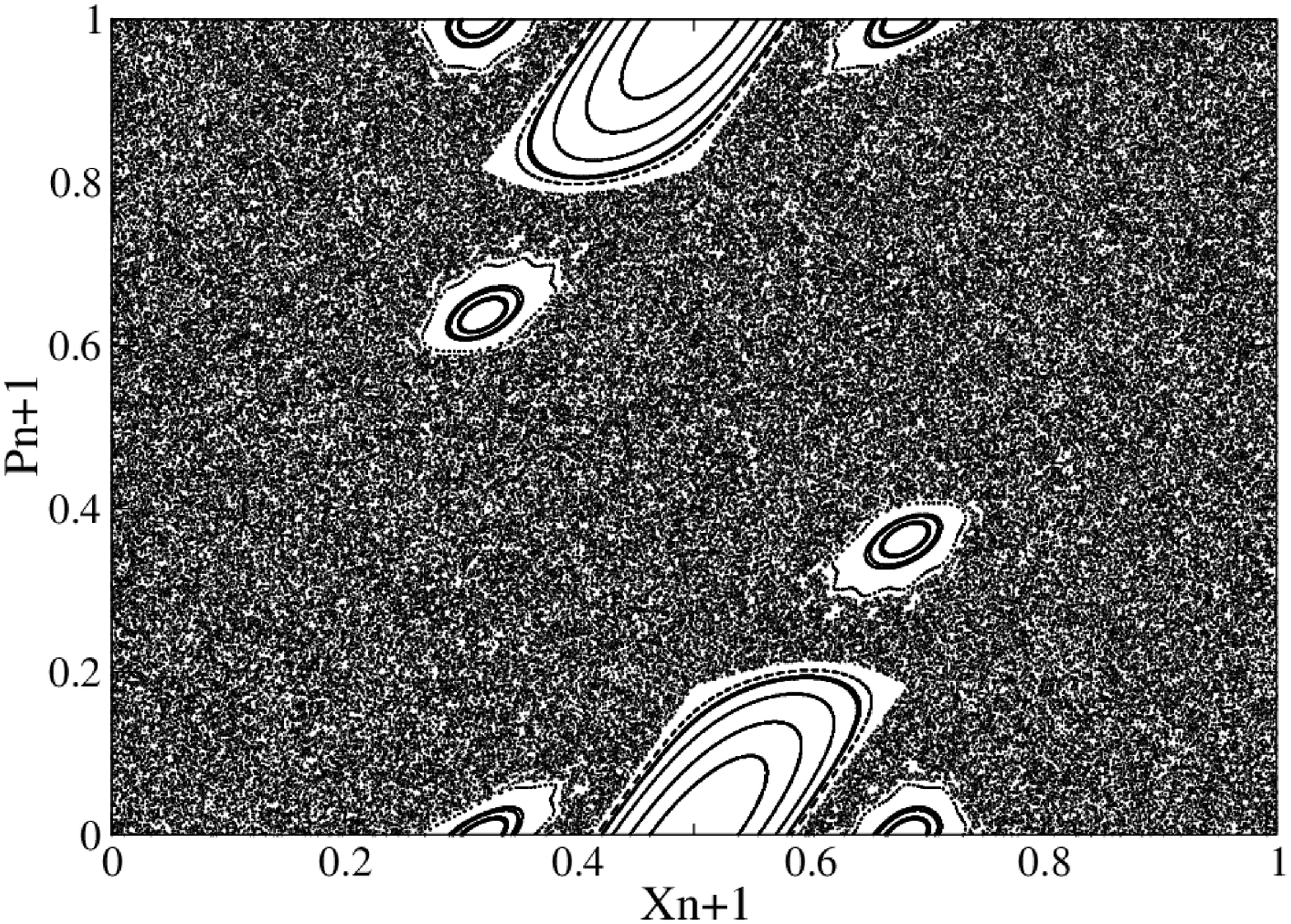}
\caption{\footnotesize{Retrato de fase do mapa padrão para $K=0,4$.}}
\end{center}
\label{rf04}
\end{figure} 
%
%
%%%%
\begin{figure}[ht!]
\begin{center}
\includegraphics[height=6cm,clip]{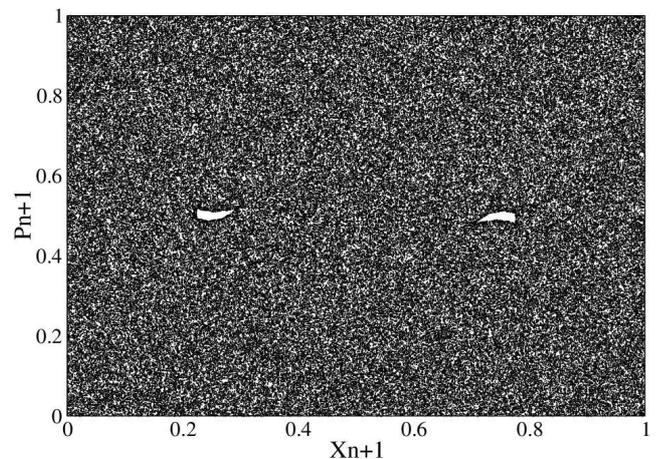}
\caption{\footnotesize{Retrato de fase do mapa padrão para $K=0,8$.}}
\end{center}
\label{rf08}
\end{figure} 

A figura (9) apresenta o retrato de fase para o Mapa Padrão com $K=0,1$. 
Nesse valor de $K$ o espaço de fases é deformado. 
Note que o sistema não é mais integrável, mas misto, com comportamento regular
e caótico coexistindo no mesmo espaço de fases. A área do regime caótico ainda 
é pequena, mas pode ser vista entre $x_{n+1}=0,0$ e $x_{n+1}=0,1$, 
$p_{n+1}=0,0$ e $p_{n+1}=0,05$, na figura (9).

Verifica-se o aparecimento de linhas fechadas, geralmente chamadas de ilhas
ou armadilhas dinâmicas. Estas ilhas rodeiam pontos, que normalmente não são 
visíveis, de equilíbrio estável, chamados de pontos Elípticos 
\cite{monteiro, lichtenberg, her2}. 
Quando se fornece para o mapa os valores 
$(x_{1}, p_{1})$ do ponto de equilíbrio $(x^{\ast}, p^{\ast})$, 
o sistema retorna sempre aos mesmos pontos nas iteração subsequêntes, 
ou seja, a dinâmica cessa e entra em equilíbrio estático com
\begin{equation}
x_{n+1}=x_{n}=x^{\ast} \qquad e \qquad p_{n+1}=p_{n}=p^{\ast}
\label{derxp}
\end{equation}

Na figura (10) pode-se ver uma grande mudança no retrato de fase do mapa, quando
o parâmetro $K$ assume o valor $0,2$. 
Com o aumento do parâmetro $K$ percebe-se que muitas linhas desaparecem
e dão lugar à pontos aleatórios, ou seja, o chamado mar caótico.
O desaparecimento das linhas deve-se a quebra da integrabilidade que é
proporcionada pela geração de frequências irracionais de rotações em torno do 
toro que forma o espaço de fases do sistema. 
A superfície desse toro é delimitada pela energia do sistema, 
que nesse caso é constante por se tratar de um sistema conservativo 
\cite{goldstein, monteiro, her2}. 

Diversos pontos elípticos ainda aparecem na 
figura (10), mostrando que dependendo das condições iniciais 
o sistema pode ser regular ou caótico. 

No entorno das ilhas, na regiõa caótica, ainda pode ser notado uma
densidade maior de pontos. Esses pontos representam o 
fenômeno chamado de {\it Stickiness} e tornam a dinâmica do sistema quase 
regular. Esse fenômeno é bem descrito na referência \cite{her2, her}.

Com $K=0,4$, a figura (11) apresenta uma diminuição acentuada das ilhas e dos 
pontos de equilíbrio. Visivelmente o retrato de fase apresenta oito pontos 
elípticos e a região caótica aumenta consideravelmente. Diz-se que o 
comportamento do sistema tornou-se mais caótico.

A figura (12) apresenta o comportamento do mapa para $K=0,8$. 
Pode-se ver que apenas duas regiões não são visitadas por pontos. 
Nestas regiões encontram-se pontos
de equilíbrio circundados por ilhas. O retrato de fase está quase totalmente 
tomado pelo caos. Se $K$ atinge o valor igual a $1$ todas as ilhas desaparecem 
e o sistema torna-se totalmente caótico ou ergódico (ver ref. 
\cite{lichtenberg}), fazendo com que todas as ilhas desapareçam.
\section{Aprendizagem Significativa}
\label{as}
As teorias de aprendizagem evoluiram para um entendimento de que o estudante 
deve fazer conexões do novo conhecimento com o que já está presente 
em sua extrutura cognitiva (conhecimento prévio). 
Esse é o caso da Aprendizagem Significativa apresentada por Ausubel 
\cite{ausubel}, e que vem sendo amplamente aplicada ao ensino de física 
\cite{moreira}.

Aprendizagem Significativa é aquela em que um novo conceito se associa 
com os outro conceitos prévio que o estudante já possui.
Para isso, o estudante utiliza uma ideia básica, âncora, que é
chamada de subsunçor. Em outras palavras o subsunçor é um conhecimento 
específico pertencente à 
estrutura cognitiva do estudante, que será associado aos novos conceitos
apresentados \`a ele.

Uma aula que utiliza a programação em Fortran como ferramenta para entender
os conceitos da física é uma aula que cria um ambiente de aprendizagem 
significativa. O ambiente de aprendizagem deixa o aluno estimulado à entender 
melhor os conceitos físicos, à conseguir relacioná-los com 
outro assunto, como o da programação, e de aprender uma linguagem de 
programação. A programação em Fortran ainda pode ser utilizada pelo estudante
para outras finalidades, como em matemática, engenharia ou aprender com mais 
facilidade outra linguagem de programação, como C$++$ ou Java. 

O aprendizado significativo ocorre de maneira que o estudante associa os 
conceitos da física aos conceitos da programação e de seus resultados.
Isso pode acontecer por que o estudante precisará utilizar todo seu conhecimento
prévio ou de física ou de programação, como base de sustentação, para 
construir um código que expresse o fenômeno físico estudado.

Podemos citar o caso das energias cinética,
potencial e mecânica no exemplo do sistema massa-mola. 
Com o entendimento prévio do conceito de energia o estudante constroi o 
código em Fortran. 
Os subsunçores são as energias e o novo conceito é a lógica da programação. 
O fato mais interessante nesse caso é do estudante poder ter mais de um 
subsunçor para o mesmo aprendizado. 
Inicialmente o subsunçor é o conceito físico, da energia. 
Após a implementação do código e de ter obtido os resultados, tem-se uma nova 
interpretação dos gráficos através da associação feita entre este conceito
e o código construído, ou seja, na forma que este o estudante cronstruíu sua
ideia aplicada ao código. Dessa forma, o novo 
subsunçor é o código, que será associado a nova interpretação do conceito 
de energia.

Neste caso, uma aprendizagem cíclica, que agrega conhecimento físico e de 
programação. Isso mostra como a ferramenta da programação é importante para 
o entendimento de forma significativa dos fenômenos físicos.

\section{Conclusão}
\label{concl}
Nos dias de hoje a programação numérica é imprescindível para a Física, 
Matemática, Engenharia e até áreas menos científicas como o cinema. Neste 
aspecto, alunos dessas áreas acadêmicas precisam de referências bibliográficas
para iniciarem seus estudos em programação. 
Neste trabalho apresentamos uma breve introdução à programação em Fortran 77, 
no intuito de oferecer uma primeira abordagem dessa linguagem com aplicações 
na Física e Matemática.

O comportamento de algumas funções trigonométricas, como o seno, são 
apresentadas para o melhor entendimento do conceito de função. 
São apresentados códigos, no programa Fortran 77,
de soma simples aplicados à mecânica Newtoniana. 
Dessa forma, o estudante pode absorver mais facilmente os conceitos envolvidos 
na Física e na programação através da construção do códigos e visualização 
gráfica dos resultados.
Deduzimos em detalhes a descrição dinâmica de uma haste em rotação, chamado de 
rotor pulsado, que recai num sistema amplamente estudado em dinâmica não-linear,
o Mapa Padrão. Através do retrato de fase desse mapa podendo visualizar 
o comportamento dinâmico que o sistema apresenta com a variação de um 
parâmetro de não-linearidade. 
A dinâmica varia de integrável para misto e, finalmente, totalmente caótico.
 
Uma breve discussão sobre a aprendizagem significativa foi apresentada com 
aplicação prática nos exemplos construídos ao longo do trabalho. 
Pode-se inferir que apresentar a linguagem de programação em Fortran 77 para 
os estudantes em sala de aula, no intuito de melhorar o entendimento da física,
pode propiciar um ambiente de aprendizagem mais significativo.

\section{Agradecimentos}
\label{agra}
Os autores agradecem o suporte financeiro da Fundação Araucária e 
do CNPq.

\end{document}